\documentstyle[12pt]{article}
\newcommand{\bee}{\begin{equation}}
\newcommand{\ene}{\end{equation}}
\newcommand{\pr}{\partial}
\newcommand{\su}{\over}
\newcommand{\al}{\alpha}
\newcommand{\px}{\phantom{i}} 
\newcommand{\pdx}{\pr\px}
\newcommand{\th}{\theta}
\begin{document}
\thispagestyle{empty}
\baselineskip6mm
\title{\vspace{-3cm}
\bf Multimomentum Maps in General Relativity}
\author{Cosimo Stornaiolo$^{1,2}$,\thanks{Electronic address:
cosmo@napoli.infn.it} \ Giampiero Esposito$^{1,2}$\thanks
{Electronic address: esposito@napoli.infn.it}}
\date{}
\maketitle
\hspace{-6mm}$^{1}${\em INFN, Sezione di Napoli,
Mostra d'Oltremare Padiglione 20, 80125 Napoli, Italy}\\ $^{2}${\em
Dipartimento di Scienze Fisiche, Mostra d'Oltremare Padiglione 19,
80125 Napoli, Italy}

\begin{abstract}
The properties of multimomentum maps on null hypersurfaces,
and their relation with the constraint analysis of
General Relativity, are described. Unlike the case of
spacelike hypersurfaces, some constraints which are
second class in the Hamiltonian formalism turn out to
contribute to the multimomentum map. 
\end{abstract}
\maketitle

\section{Introduction}
\hspace{\parindent}
In order to quantize gravity, a very long-time effort has been produced by
physicists over the last fifty years. Since the perturbative approach
fails to produce a renormalizable theory, it seemed more viable to
proceed with the analysis of canonical gravity, 
which leads to a non-perturbative approach
to quantum gravity. The canonical quantization of
field theories follows Dirac's prescription to translate the Poisson
brackets among the canonical variables into commutators of the operators
corresponding to these variables, and a special treatment is reserved to
those systems and fields which are constrained. 
Within this framework Dirac, Bergmann, Arnowitt-Deser-Misner, Isham
and, more recently, Ashtekar, pursued the aim of building a
canonical formalism for General Relativity. 

The canonical approach has been successful, but it faces two important
problems. First, to obtain a Hamiltonian formulation it is
necessary to break manifest covariance. The other problem is that one
has to deal with an infinite number of degrees of freedom when a field
theory is considered. 

In recent work [1], the authors have studied a
multisymplectic version of General Relativity. In this approach, 
field theories can be treated as
an extension of the usual symplectic treatment of classical mechanics. 
Here, instead of working with an infinite number of degrees of 
freedom, as is usually done with the symplectic 
approach, the whole theory is
constructed on a 1-jet bundle, whose local coordinates are  
spacetime coordinates, the fields and their first derivatives.

It has been shown, in particular, 
how the classical multisymplectic analysis of the 
constraints is equivalent to the constraint analysis given by Ashtekar 
from a canonical point of view [1]. 
Since the constraints were studied on a spacelike hypersurface, in this
paper, to complete our previous investigations,
null hypersurfaces are considered. In the
Hamiltonian formulation of General Relativity, the constraint analysis
on null hypersurfaces plays an important role since such surfaces
provide a natural framework for the study of gravitational radiation in
asymptotically flat space-times [2-7]. Moreover, in a null canonical
formalism, the physical degrees of freedom and the observables of the
theory may be picked out more easily [5,6]. 
Therefore it appears very interesting  
to extend the constraint analysis of
[1] to null hypersurfaces and find out 
whether equivalent results exist. This may also provide
further insight into the techniques for dealing
with second-class constraints [8].

In section 2, null tetrads are defined according to 
[5]. In section 3, constraints are studied 
for a self-dual action.
Concluding remarks are presented in section 4.
\section{Null Tetrads}
\hspace{\parindent}
In this paper we are only concerned with 
the local treatment of the problem on 
null hypersurfaces. Thus, many problems arising 
from the possible null-cone
singularities are left aside. To obtain a consistent 3+1
description of all the Einstein equations, 
one can introduce the null tetrad [5]
\bee
e_{\hat 0}={1\over N}\left({\pdx\over\pr t} 
- N^i{\pdx\su\pr x^i}\right)
\ene
and
\bee
e_{\hat k}=\left(v_{\hat k}^i 
+ \al_{\hat k} {N^{i}\over N} \right){\pdx\su\pr x^i}-  
{\al_{\hat k}\over N} {\pdx\su\pr t} \; .
\ene
The duals to these vectors are
\bee
\th^{\hat 0}=(N+\al_i N^i)dt + \al_i dx^i
\ene
and
\bee
\th^{\hat k}=\nu^{\hat k}_{i}(N^{i}dt+ dx^{i}) \; ,
\ene
where
\bee
v_{\hat k}^i\nu^{\hat l}_i=\delta^{\hat l}_{\hat k} 
\ene
and
\bee
\al_{\hat k}=v^i_{\hat k}\al_{i} \; .
\ene
The Minkowski metric is given by
\bee
\eta_{\hat a\hat b}=\eta^{\hat a\hat b}=
\left(\matrix{0&1&0&0\cr
        1&0&0&0\cr 
	0&0&0&-1\cr
	0&0&-1&0\cr}\right) \; .
\ene
The tetrad labels $\hat a$, $\hat b$, $\hat c$ range
from $0$ through $3$, while the indices 
$\hat k$, $\hat l$ range from $1$ through $3$.
Analogous notation is used for the spacetime indices $a$,$b$,... and
$i$, $j$, etc.

It is straightforward to see that 
\bee
g^{ab}t_{,a}t_{,b}=-{2\su
N^2}\,(\al_{\hat 1} + \al_{\hat 2} \al_{\hat 3}) \; . 
\ene
This implies that
the hypersurfaces $t=$const. 
are null if and only if $\al_{\hat 1} +
\al_{\hat 2} \al_{\hat 3}=0$. By a particular choice of coordinates, it is
always possible to set $ \al_{\hat 2}= \al_{\hat 3}=0 $.

In many of the following equations the tetrad vectors appear in the 
combination
\bee
{\tilde p}^{\,ac}_{\phantom{\,ac} \hat a\hat c } = 
{e\over 2} \left(e^a_{\hat a}
e^{c}_{\hat c} - e^a_{\hat c}e^{c}_{\hat a} \right) \; ,
\ene
where $e=N\nu$ with $\nu=det(\nu^{\hat a}_i)$.
\section{The Self-Dual Action}
\hspace{\parindent}
In [5,7] the Hamiltonian formulation of a complex self-dual action
on a null hypersurface in Lorentzian space-time 
was studied. The 3+1 decomposition was inserted into
the Lagrangian, and the constraints were derived with the usual
Dirac's procedure. In this section the 
results of [5] are briefly
summarized and then compared with the 
corresponding constraints obtained by
the multimomentum map. Since these constraints correspond
to the secondary constraints of the
Hamiltonian formalism [8], the discussion is focused on them.
 
The complex self-dual part of the connection are the complex 
one-forms given by
\bee
{ }^{(+)}\omega_a^{\phantom{a}\hat a\hat c} =
{1\over2}\left( \omega^{\phantom{a}\hat a\hat c}_a - {i\over 2}
\epsilon^{\hat a\hat c}_{\phantom{ac} \hat b\hat d}\,
\omega_a^{\phantom{a}\hat b\hat d} \right) \; .
\ene
Explicitly, one has
\bee
{ }^{(+)}\omega^{\phantom{a}\hat 0\hat 1 }_{a}
={ }^{(+)}\omega^{\phantom{a}\hat 2\hat 
3 }_{a}={1\over2}\left( 
\omega^{\phantom{a} \hat 0\hat 1}_{a} 
+ \omega^{\phantom{a} \hat 2\hat 3}_{a}
\right) \; ,
\ene
\begin{eqnarray}
{ }^{(+)}\omega^{\phantom{a}\hat 2\hat 1 }_{a}
&=& 
\omega^{\phantom{a} \hat 2\hat 1}_{a} \; , \;
{ }^{(+)}\omega^{\phantom{a}\hat 0\hat 3 }_{a} = 
\omega^{\phantom{a} \hat 0\hat 3}_{a} \; , \;
\nonumber\\
&\;& { }^{(+)}\omega^{\phantom{a} \hat 0\hat 2}_{a}=
{ }^{(+)}\omega^{\phantom{a}\hat 1\hat 3}_{a}=0 \; .
\end{eqnarray}
The curvature of a self-dual connection is equal to the self-dual 
part of the curvature:
\bee
\Omega({ }^{(+)}\omega)={ }^{(+)}\Omega(\omega) \; .
\ene
Thus, the complex self-dual action to be considered is [1]
\bee
S_{SD}\equiv {1\over2}\int_M d^4x\,e\,e^a_{\hat a} \,e^b_{\hat b} \,
{ }^{(+)}\Omega^{\phantom{ab}\hat a\hat b}_{ab} \; .
\ene
The tetrad vectors occur in the 
following equations in a combination 
which is the self-dual part of eq. (9).
The 13 secondary constraints obtained in the Hamiltonian formalism in
[5] after the 3+1 split, 
and written with the notation of the
present paper, are as follows: 
\begin{eqnarray}
{\cal H}_{0} &\equiv& 
-\left({e\over N}\right)^{2} V_{\hat 2}^{i}
\Bigr[{ }^{(+)}\Omega_{ij}^{\phantom{ij}\hat 
0\hat 1}
V_{\hat 3}^{j} \nonumber\\
&+& { }^{(+)}\Omega_{ij}^{\phantom{ij}
\hat 2\hat 1}V_{\hat 1}^{j}\Bigr]
\approx 0 \; ,
\end{eqnarray}
\bee
{\cal H}_{i} \equiv {e\over N}\left[
{ }^{(+)}\Omega_{ij}^{\phantom{ij}\hat 0\hat 1}V_{\hat 1}^{j}+
{ }^{(+)}\Omega_{ij}^{\phantom{ij}\hat 0\hat 3}V_{\hat 3}^{j}
\right] \approx 0 \; ,
\ene
\bee
{\cal G}_{\hat 1} \equiv
-\pr_{i}\left({e\over N}V^{i}_{\hat 1}\right) 
-2{e\over N} { }^{(+)}\omega_{i}^{\; \; {\hat 0}{\hat 3}} 
V_{\hat 3}^{i} \approx 0 \; ,
\ene
\bee
{\cal G}_{\hat 2}\equiv
-{e\over N}{ }^{(+)}\omega_{i}^{\phantom{i}
\hat 0\hat 3}V^{i}_{\hat 1}\approx 0 \; , 
\ene
\begin{eqnarray}
{\cal G}_{\hat 3}&\equiv&
-\pr_{i}\left({e\over N}V^{i}_{\hat 3}\right) 
+{e\over N} { }^{(+)}\omega_{i}^{\phantom{i}
\hat 2\hat 1}V_{\hat 1}^{i} \nonumber\\
&+& 2{e\over N}{ }^{(+)}\omega_{i}^{\phantom{i}\hat 0\hat 1}
V^{i}_{\hat 3}\approx 0 \; ,
\end{eqnarray}
\begin{eqnarray}
\chi^{i}&=& -2 \partial_{j}{e^{2}\over N}
V_{{\hat 2}}^{\; [i} \; V_{{\hat 1}}^{\; j]}
\nonumber \\
&-& 2{e^{2}\over N} 
{ }^{(+)}\omega_{j}^{\; {\hat 0}{\hat 3}}
V_{{\hat 2}}^{\; [i} \; V_{{\hat 3}}^{\; j]}
\nonumber \\
&-& 4{e^{2}\over N} { }^{(+)}\omega_{j}^{\; {\hat 0}{\hat 1}}
\; V_{{\hat 2}}^{\; [i} \; V_{{\hat 1}}^{\; j]}
\nonumber \\
&+& 2 {e\over N} { }^{(+)}\omega_{j}^{{\hat 0}{\hat 3}}
\; N^{[i} \; V_{{\hat 1}}^{\; j]}
\nonumber \\
&+& {e\over N} { }^{(+)}\omega_{0}^{\; {\hat 0}{\hat 3}}
\; V_{{\hat 1}}^{\; i}
\approx 0 \; ,
\end{eqnarray}
\bee
\phi_{i}\equiv -{e\over N} \left[
{ }^{(+)}\Omega^{\phantom{ij}\hat 0\hat 1}_{ij} V_{\hat 3}^{j} + 
{ }^{(+)}\Omega^{\phantom{ij}\hat 2\hat 1}_{ij} V_{\hat 1}^{j}
\right] \approx 0 \; .
\ene
The irreducible second-class constraints turn out to be
${\cal H}_{0},{\cal G}_{\hat 3},\chi^{i},\phi_{i}V_{\hat 2}^{i}$
and $\phi_{i}V_{\hat 3}^{i}$ [5]. Note that, following [5,7],
we have set to zero all the $\alpha$ parameters in the course
of deriving eqs. (15)--(21).
 
Let us now discuss the constraints from our point of
view. The multimomentum map is [9]
\begin{eqnarray}
I_{{\cal S}_N}^{+}[\xi,\lambda]&=&\int_{{\cal S}_N}\biggr[
{ }^{(+)}{\widetilde p}_{\; \; \; \; {\hat b}{\hat d}}^{\; ac}
\biggr(\xi_{\; \; ,a}^{b}
\; { }^{(+)}\omega_{b}^{\; \; {\hat b}{\hat d}} \nonumber\\
&-&(D_{a} { }^{(+)}\lambda)^{{\hat b}{\hat d}}
+{ }^{(+)}\omega_{a \; \; \; \; ,b}^{\; \; {\hat b}{\hat d}}
\; \xi^{b}\biggr) \nonumber \\
&+& {1\over 2} 
{ }^{(+)}{\widetilde p}_{\; \; \; \; {\hat b}{\hat d}}^{\; ab}
\; { }^{(+)}\Omega_{ab}^{\; \; \; {\hat b}{\hat d}}
\; \xi^{c}\biggr]d^{3}x_{c} \; .
\end{eqnarray}
The constraint equations obtained from setting to zero 
this multimomentum map are then
\begin{eqnarray}
&\;& \int_{{\cal S}_N} { }^{(+)}\lambda^{{\hat 0}{\hat 1}}
\biggr[\pr_{i}
\Bigr({e\over N}V_{\hat 1}^{i}\Bigr) \nonumber\\
&+& {e\over N}
{ }^{(+)}\omega_{i \hat 1}^{\phantom{i \hat 1}\hat 3}V_{\hat 3}^{i}
\biggr]d^{3}x_{0} =0 \; ,
\end{eqnarray}
\begin{eqnarray}
&\;& \int_{{\cal S}_N} { }^{(+)}\lambda^{{\hat 0}{\hat 3}}
\biggr[\pr_{i}
\Bigr({e\over N}V_{\hat 3}^{i}\Bigr) \nonumber\\
&+& {e\over N}
{ }^{(+)}\omega_{i \hat 3}^{\phantom{i \hat 3}\hat l}V_{\hat l}^{i}
\biggr]d^{3}x_{0} =0 \; , 
\end{eqnarray}
\bee
\int_{{\cal S}_{N}} { }^{(+)}\lambda^{{\hat 1}{\hat 2}} D_{i}
{ }^{(+)}{\widetilde p}_{\; \; \; {\hat 1}{\hat 2}}^{\; 0i}
\; d^{3}x_{0}=0 \; ,
\ene
\begin{eqnarray}
&\;& \int_{{\cal S}_{N}}  
e V_{\hat 2}^{i}\left[ 
{ }^{(+)}\Omega^{\phantom{ij} \hat 0\hat 1}_{ij} 
V_{\hat 3}^{j}
+{ }^{(+)}\Omega^{\phantom{ij} \hat 
2\hat 1}_{ij} V_{\hat 1}^{j}\right]\xi^{0}d^{3}x_{0} \nonumber\\
&-& \int_{{\cal S}_{N}} 
{2e\over N}N^{i} 
\Bigr[{ }^{(+)}\Omega^{\phantom{ij} \hat 0\hat 1}_{ij}
V_{\hat 1}^{j} \nonumber\\
&+& { }^{(+)}\Omega^{\phantom{ij} \hat 0\hat 3}_{ij} 
V_{\hat 3}^{j}
\Bigr] \xi^{0} \, d^{3}x_{0} = 0 \; ,
\end{eqnarray}
\begin{eqnarray}
&\;& \int_{{\cal S}_N} 
{e\over N} \biggr[
{ }^{(+)}\Omega^{\phantom{ij} \hat 0\hat 1}_{ij} 
V_{\hat 1}^{i} \nonumber\\
&+& { }^{(+)}\Omega^{\phantom{ij} \hat 0\hat 3}_{ij} 
V_{\hat 3}^{i}
\biggr]\xi^{j}\, d^{3}x_{0} = 0 \; .
\end{eqnarray}

On the other hand,
the Euler-Lagrange equations resulting from the action (14) are
\bee
e^{b}_{\;\hat b}\left[{ }^{(+)}\Omega^{\phantom{bh}\hat b\hat c}_{bh}-
{1\over2}e^{d}_{\;\hat a}e^{\hat c}_{\; h}
{ }^{(+)}\Omega^{\phantom{bd}\hat b\hat a}_{bd}\right]=0 \; ,
\ene
and
\bee
D_{b}{ }^{(+)}{\widetilde p}^{\;\, ab}_{\phantom{\;\,ab} 
\hat a \hat b}=0 \; .
\ene
The self-dual Einstein equations (28) can be written 
explicitly in the form
\bee
{ }^{(+)}G^{\hat 0}_{h} \equiv 
e^{b}_{\hat 1} \; 
{ }^{(+)}\Omega_{bh}^{\phantom{bh}\hat 1\hat 0} +
e^{b}_{\hat 3} \; 
{ }^{(+)}\Omega_{bh}^{\phantom{bh}\hat 3\hat 0} = 0 \; ,
\ene
\bee
{ }^{(+)}G^{\hat 1}_{h} \equiv
e^{b}_{\hat 0} \; 
{ }^{(+)}\Omega_{bh}^{\phantom{bh}\hat 0\hat 1} +
e^{b}_{\hat 2} \; 
{ }^{(+)}\Omega_{bh}^{\phantom{bh}\hat 2\hat 1} = 0 \; ,
\ene
\bee
{ }^{(+)}G^{\hat 2}_{h} \equiv
e^{b}_{\hat 1} \; 
{ }^{(+)}\Omega_{bh}^{\phantom{bh}\hat 1\hat 2} +
e^{b}_{\hat 3} \; 
{ }^{(+)}\Omega_{bh}^{\phantom{bh}\hat 3\hat 2} = 0 \; ,
\ene
\bee
{ }^{(+)}G^{\hat 3}_{h} \equiv
e^{b}_{\hat 0} \; 
{ }^{(+)}\Omega_{bh}^{\phantom{bh}\hat 0\hat 3} +
e^{b}_{\hat 2} \; 
{ }^{(+)}\Omega_{bh}^{\phantom{bh}\hat 2\hat 3} = 0 \; .
\ene
It is easy to show that the equations independent of time
derivatives on a null hypersurface are the spatial components of eqs. 
(30) and (32), jointly with the equations
\bee
D_{i}{ }^{(+)}{\widetilde p}_{\; \; \; {\hat 0}
{\hat 1}}^{\; 0i}
=D_{i}{ }^{(+)}{\widetilde p}_{\; \; \; {\hat 0}
{\hat 3}}^{\; 0i}=0 \; ,
\ene
\bee
D_{i}{ }^{(+)}{\widetilde p}_{\; \; \; {\hat 1}
{\hat 2}}^{\; 0i}=0 \; ,
\ene
which are equivalent to (23)--(25), and 
\bee
D_{j}{ }^{(+)}{\widetilde p}_{\; \; \; {\hat 1}
{\hat 2}}^{\; ij}=0 \; .
\ene
The comparison of eqs. (15)--(21) with eqs. (23)--(27) shows
that eq. (23) corresponds to eq. (17), 
eq. (24) to eq. (19), eq. (25) to eq. (18),
eq. (26) to eqs. (15) and (16), and eq. (27) 
to eq. (16). 

Remarkably, the second-class constraints ${\cal H}_{0}$
and ${\cal G}_{{\hat 3}}$ are found to contribute to
the multimomentum map (see section 4).
\section{Concluding Remarks}
\hspace{\parindent}
This paper has considered the application of the multimomentum-map
technique to study General Relativity as a constrained
system on null hypersurfaces. Its contribution lies in relating
different formalisms for such a constraint analysis.
We have found that, on null hypersurfaces, the multimomentum
map provides just a subset of the full set of constraints of the
theory, while the other constraints turn out to be those particular
Euler-Lagrange equations which are not of evolutionary type [9].  

Although the multimomentum map is expected to yield only 
the secondary first-class constraints [8], we have found
that some of the constraints which are second class in 
the Hamiltonian formalism occur also in the multimomentum map.
The group-theoretical interpretation of this property 
seems to be that our analysis remains covariant in that 
it deals with the full diffeomorphism group of spacetime,
say $Diff(M)$, jointly with the internal rotation
group $O(3,1)$. Hence one incorporates some constraints
which are instead ruled out if one breaks covariance, which
amounts to taking subgroups of the ones just mentioned.
In other words, only when $Diff(M)$ and $O(3,1)$ are 
replaced by their subgroups 
$Diff({\cal S}_{N}) \times Diff(\Re)$ and
$O(3)$, the constraints (24) and (26) become second class and 
hence do not contribute to the multimomentum map.

In [10], the holomorphic multimomentum map has been studied to
obtain a formulation of complex general relativity, and this
appears to be another challenging field of research.

The authors are indebted to Mark Gotay and Luca Lusanna
for very useful discussions.

\end{document}